\documentclass[twocolumn,pra,amsmath,amssymb,showpacs]{revtex4}
\usepackage[utf8]{inputenc}
\usepackage{amsmath}
\usepackage{amsfonts}
\usepackage{amssymb}
\usepackage{color}

\begin{document}

\title{Quantum discord and remote state preparation}

\author{Gian Luca Giorgi}
\affiliation{INRIM, Strada delle Cacce 91, I-10135 Torino, Italy }

\begin{abstract}
The role played by quantum discord in mixed-state computation is widely debated since, in spite of evidence of its importance in creating quantum advantages, even in the absence of entanglement, there are not direct proofs of its necessity in these computational tasks. Recently the presence of discord was shown to be  necessary and sufficient  for remote state preparation for a broad class of quantum channels [B. Daki\'c \textit{et al.}, Nat. Phys. \textbf{8}, 666 (2012)]. Here, we show that this property is not universal. There are states whose discord cannot be considered as a quantum resource, since it has been produced locally, that are useful for remote state preparation, and there are bona fide discordant states that are of no help.

\end{abstract}


\maketitle

\section{Introduction}
Quantifying and characterizing the nature of correlations in a quantum state, besides the fundamental scientific interest, has a crucial applicative importance for the full development of quantum technologies \cite{horodecki}. In the early stage of the field of quantum information and computation, only tasks involving pure states were considered. In that scenario, as proven in Ref. \cite{linden}, exponential computational speedup is possible only if entanglement grows with the size of the system. Therefore, entanglement was identified as the unique quantum-mechanical trait.
On the other hand, once mixed states are taken into account for computational purposes, the role played by entanglement becomes less clear. For instance, in the
so-called deterministic quantum computation with one qubit
(DQC1) protocol \cite{knill} the estimation of the normalized trace of a unitary matrix of dimension $n$ can be attained in a number of trials that does not scale exponentially with $n$. In this protocol, the speedup can be achieved using
factorized states. As suggested in Ref. \cite{datta}, speedup could be due
to the presence of another quantifier, the so called quantum
discord, whose definition originates from the observation that there exist two quantum analogs of the classical mutual information \cite{zurek,vedral,rmp}, since this measure of correlations is present in the final stage of the computation.

There is no clear evidence of the relation between entanglement and quantum discord, since they seem
to capture different properties of the states. They coincide for pure states but can be substantially different if statistical mixtures are taken into account \cite{mdms}. A discussion about the interplay between these two quantities can be found in Refs. \cite{cornelio,streltsov,piani}.
 In contrast to entanglement, discord can be generated using local noise \cite{noisediscord} and  is not monogamous \cite{monogamy}.

Apart from the DQC1 protocol, there is other evidence of quantum advantage generated in the absence of entanglement. For instance, it was shown that the presence of discord is essential in quantum state discrimination \cite{qsd}.  
 More recently, in Ref. \cite{dakic}, Daki\'c {\it et al.} showed, both theoretically and experimentally, that quantum discord is a resource for remote state preparation (RSP). The result was obtained considering a broad class of two-qubit states. For this family of states, it was proven that the RSP fidelity ${\cal F}$ is equal to the square of a geometric version of  discord \cite{dakic2}, providing an operational meaning for this quantity. A different operational meaning of quantum discord in terms of state
merging was proposed in Refs. \cite{opmean}.

Based on these observations, even if its role is still widely debated, quantum discord has been identified as the quantifier of the amount of quantum correlations of physical states. However, any good measure of correlations between two (or more)  parties should satisfy a series of reasonable and widely accepted  properties. One of these properties states that correlations must be nonincreasing under local operations. In fact, as shown in Refs. \cite{noisediscord}, unlike mutual information and classical correlations, quantum discord does violate this rule. In other words, given a classical state,  it is sometimes possible to generate discord by applying local noise only. Then, as suggested by  Gessner \textit{et al.} \cite{gessner}, a more accurate indicator of quantum correlations should be identified,  the discord by itself being unable to discriminate between correlations that are bona fide nonlocal and ``spurious" correlations that can be created locally.
A witness able to identify which states are really quantum and which are not, according to the meaning of this expression given before, is given by the so-called ``correlation rank" \cite{dakic,gessner}  that quantifies the minimal amount of bipartite product operators needed in a linear combination to represent a given quantum state.

The main scope of this paper is to discuss the role played by  quantum correlations  in the RSP task described in Ref. \cite{dakic}, paying special attention to those  states whose discord has been created by locally acting on  classical states. It will be proven that dealing with bona fide quantum correlations is neither sufficient nor necessary for RSP. The paper is structured as follows. In Sec. \ref{II}, we review the definition of quantum discord and generalize it in order to exclude that part of correlations that has been created locally. In Sec. \ref{III} we recall the definition of correlation rank, which will be used in Sec. \ref{IV}, where it will be shown that the advantage due to the presence of  quantum discord can manifest itself even if that discord has been created locally. The conclusions are given in Sec. \ref{V}.

\section{Quantum correlations}\label{II}
 Given a bipartite quantum state $\varrho$, its discord can be considered as a measure of how much
disturbance is caused when trying to learn about party $A$ when measuring party $B$. It is defined as
\begin{equation}
{\cal D}_{A:B}={\cal I}(\varrho)-{\cal J}_{A:B},
\end{equation}
where ${\cal I}(\varrho)=S(\varrho_{A})-S(\varrho_{B})-S(\varrho)$ is the quantum version of the mutual information ($S(.)$ is the von Neumann entropy) and the classical correlations  are defined as
\begin{equation}
 {\cal J}_{A:B}= \max_{\{E_j^B\}}[S(\varrho_A)-S(A|\{E_j^B\})],\label{clas}
\end{equation}
 with the conditional entropy given by $S(A|\{E_j^B\})=\sum_j p_j S(\varrho_{A|E_j^B})$, $p_j={\rm
Tr}_{AB}(E_j^B\varrho)$ and where $\varrho_{A|E_j^B}= E_j^B\varrho /{p_j} $ is the density
matrix after a positive operator valued measure (POVM) $\{E_j^B\}$ has been performed on $B$. 

A slightly different definition of discord, based on geometric considerations, was given in Ref. \cite{dakic2}. There, discord was defined as the distance, as measured by the square norm in the Hilbert-Schmidt space, between the density matrix under study and the closest element belonging to the set of zero-discord states. 

 The easiest way to wash out from ${\cal D}_{A:B}$  the amount of correlation that could have been induced by local noise is to take into account any possible state  $\varrho^\prime$ such that there exists a local channel $\cal M$ mapping such a state onto the target state $\varrho$:
\begin{equation}
\varrho=\frac{M_A\otimes M_B \varrho^\prime M_A^\dag\otimes M_B^\dag}{{\rm Tr} (M_A\otimes M_B \varrho^\prime M_A^\dag\otimes M_B^\dag)}.\label{map}
\end{equation} 
 Then  quantum correlations ${\cal Q}_{A:B}$ can be defined as the minimum quantum discord present in the family $ \varrho^\prime$ compatible with Eq. (\ref{map}), that is
 \begin{equation}
 {\cal Q}_{A:B}=\min_{\varrho^\prime\in {\cal M}} {\cal D}_{A:B}.\label{Q}
 \end{equation}
 As a  tautological consequence of this definition, $ {\cal Q}_{A:B}$ is not increasing under local operations and ${\cal Q}_{A:B}\le {\cal D}_{A:B}$. 
 
While the explicit calculation of   ${\cal Q}_{A:B}$ will generally represent a formidably complicated task, adding the minimization over all possible noisy local histories to the measurement minimization  appearing in the definition of quantum discord, the presence of of nonvanishing  ${\cal Q}$, as explained in next section, can be witnessed by monitoring the correlation rank \cite{dakic2,gessner}.

%
\section{Correlation rank}\label{III}
 Let us briefly recall the definition and the main properties of the  correlation rank, which will be used afterwards.  Given a bipartite Hilbert space ${\cal H}_A \otimes  {\cal H}_B$, a state $\rho$ can be decomposed as a sum of arbitrary bases of Hermitian operators $\{A_i\}$ and $\{B_i\}$. We can write 
\begin{equation}
 \rho=\sum_{n=1}^{d_A^2}\sum_{m=1}^{d_B^2}r_{nm}A_n\otimes B_m,
\end{equation}
where $d_A$ ($d_B$) is the dimension of the Hilbert space $A$ ($B$).
By means of this representation, we have introduced the correlation matrix $R=(r_{nm})$, which  can be rewritten using its singular value decomposition and cast in a diagonal representation as \cite{dakic}
\begin{equation}
 \rho=\sum_{n=1}^{L_R} c_nS_n\otimes F_n.
\end{equation}
Here, $L_R$ is the rank of $R$ and quantifies how many product operators are needed to represent $\rho$.

The value of $L_R$ can be used to witness the presence of quantum correlations. For classical states,  $L_R$ is bounded from above  by the minimum dimension of the subsystems   $d_{\text{min}}=\min\{d_A,d_B\}$. On the other hand, in general, the correlation rank is bounded by the square of $d_{\text{min}}$:  $L_R\leq d_{\text{min}}^2$. Therefore,  states with  $L_R>d_{\text{min}}$ will be necessarily discordant \cite{dakic}. However, as shown in Ref. \cite{gessner}, for quantum states whose discord can be created by local operations, $L_R\leq d_{\min}$. For this family of states, the value of the rank of their correlation matrix $L_R$ is compatible with the one of a classical state. According to Eq. (\ref{Q}),  if  $L_R> d_{\min}$, then ${\cal Q}>0$.

\section{Discord and remote state preparation}\label{IV}

In an RSP protocol, Alice wants to send to Bob a known pure state \cite{bennett}. To accomplish the RSP  task, Alice and Bob need to share a common channel $\rho$ that need to be correlated.
In the following, we specialize in the version of the protocol described in Ref. \cite{dakic}, where Alice wants to remotely prepare a quantum state on the equatorial plane of the Bloch sphere.
In the best case scenario, when $\rho$ is a Bell state, Bob can reconstruct the state Alice wanted to send to him with fidelity ${\cal F}=1$. On the other hand, if the channel is prepared in a mixed state, the fidelity and the efficiency of the protocol ${\cal P}=(2{\cal F}-1)^2$  will be reduced. In Ref. \cite{dakic}, Dakic \textit{et al.} gave an analytical expression of ${\cal F}$ as a function of the eigenvalues of the correlation tensor of $\rho$. Given two qubits, it is always possible to represent their state in the Bloch form:
\begin{equation}\label{bloch}
 \rho=\frac{1}{4}[\openone\otimes\openone+\sum_i x_i\sigma_i\otimes \openone+\sum_i y_i\openone\otimes\sigma_{i}+\sum_{i,j}T_{ij}\sigma_{i}\otimes\sigma_{j}],
\end{equation}
where $\openone$ is the $2\times 2$ identity matrix, $\sigma_{i}$ ($i=1,2,3$) are the three Pauli matrices, $x_i=\mathrm{Tr}[\rho(\sigma_i\otimes\openone)]$ and $y_i=\mathrm{Tr}[\rho(\openone\otimes\sigma_i )]$ are components of the local Bloch vectors $\vec x=\{x_1,x_2,x_3\}$ and $\vec y=\{y_1,y_2,y_3\}$, and $T_{ij}=\mathrm{Tr}[\rho(\sigma_i\otimes\sigma_j)]$ are the elements of the correlation tensor $T$. 
The RSP fidelity is
\begin{equation}\label{fidelity}
{\cal F}=\frac{1}{2}(T_1^2+T_2^2),
\end{equation}
where $T_1^2$ and $T_2^2$ are the two lowest eigenvalues of $T^TT$ \cite{dakic}. On the other hand, for a zero-discord state, the tensor $T$ obeys $T={\rm diag}(T_1,0,0)$, considering that correlations are present in one basis only. Then, ${\cal F}=0$ for all zero-discord states. Furthermore, for a broad class of states, ${\cal F}$ is equal to the square of the geometric discord, providing an operational meaning of the latter. States satisfying such a property are either those  with maximally mixed
marginals ($\vec{x}=\vec{y}=0$) or are isotropically correlated ($T=\lambda \openone$) \cite{dakic}.

As stated in the Introduction, there are good reasons to assume  that locally induced discord cannot be considered   a legitimate quantum resource, even if there are no protocols where this common wisdom could be verified analytically.
In the following, we shall show that ${\cal F}>0$ can also be obtained using quantum states whose quantum correlations could be the result of local noise.

For the case of two qubits, which will be discussed here, the correlation matrix of any density matrix can be obtained directly from the Bloch representation of Eq. (\ref{bloch}), since the identity and the three Pauli matrices represent a complete set of Hermitian operators. Then, dropping the irrelevant (to our scope) prefactor, we have
 \begin{equation}
R=\left(\begin{array}{cccc}
1&x_1&x_2&x_3\\y_1& T_{11}&T_{12}&T_{13}\\y_2&T_{21}&T_{22}&T_{23}\\y_3&T_{31}&T_{32}&T_{33}
\end{array}\right).
\end{equation}
We know that i) for locally induced discord $L_R=2$ and ii) for classical states $L_T<2$, where $L_T$ is the rank of the correlation tensor $T$. We also know that $L_R$ is nonincreasing under local operations \cite{gessner}.
Here, we  show that  $L_T$  can increase under local operations. Let us consider the  classically correlated  state 
\begin{equation}
 \rho_{cl}=
\frac{1}{2}\left(|00\rangle \langle 00|+|11\rangle \langle 11|\right).
\end{equation}
It is easy to verify that $R= {\rm diag}(1,0,0,1)$. Then,  $L_R=2$ and $L_T=1$.
By applying the local operation $\Phi\otimes \Phi$, defined through $\Phi(X)={|0\rangle\langle 0|X|0\rangle\langle 0|}+{|+\rangle\langle 1|X|1\rangle\langle +|}$, to both qubits, we get
\begin{equation}
 \tilde{\rho}=(\Phi\otimes \Phi)\rho_{cl}=\frac{1}{2}\left(|00\rangle \langle 00|+|++\rangle \langle ++|\right),
 \end{equation} 
where $|+\rangle=(|0\rangle+|1\rangle)/\sqrt{2}$. Its correlation matrix reads
\begin{equation}
R_{ \tilde{\rho}}=\left(\begin{array}{cccc}
1&\frac{1}{2}& 0 &\frac{1}{2}\\
\frac{1}{2}&\frac{1}{2}&0 &0\\
0&0&0&0\\\frac{1}{2}&0&0&\frac{1}{2}
\end{array}\right).
\end{equation}
As the correlation tensor $T$ is already diagonal, it is immediate to verify that its rank is now $L_T=2$. Then, we started introducing a classical state with $L_T=1$, and through local manipulations, we arrived at a new state with $L_T=2$. During the process, $L_R$ did not change, as expected. Surprisingly enough, as an effect of the increase of $L_T$, according to Eq. (\ref{fidelity},) $\tilde{\rho}$ supports a nonvanishing  RSP fidelity even if ${\cal Q} (\tilde{\rho})=0$.

We have just shown that there are ``fake" quantum states, with correlation rank $L_R=2$, that are useful for RSP. In the following, we shall prove the complementary result; that is, there are bona fide quantum states whose RSP fidelity is equal to zero. 
In fact, unfortunately, there are states with  ${\cal Q}>0$ that
are not good resources.
It is, indeed, possible to build density matrices characterized by $L_R=3$ and $L_T=1$ at the same time. The condition $T={\rm diag}(T_1,0,0)$ used to classify classical states is in fact necessary but not sufficient.  Therefore, there exist matrices that are bona fide quantum correlated but  are not useful for RSP.  A family of such states is obtained by considering physically meaningful density matrices with $\rho_{11}-\rho_{22}=\rho_{44}-\rho_{33}$ together with $\rho_{14}=\rho_{23}=0$ and with $\rho_{12}=\rho_{13}=\rho_{21}=\rho_{31}=\rho_{42}=\rho_{43}=\rho_{24}=\rho_{34}$. 
An explicit example is the following:
\begin{equation}
\sigma=\left(\begin{array}{cccc}
0.2& 0.1 &  0.1 & 0 \\
 0.1 & 0.1 & 0 & 0.1\\
0.1 & 0 & 0.3 & 0.1\\
0&  0.1 &  0.1 & 0.4
\end{array}\right).
\end{equation}
Its correlation matrix is (apart from the factor $1/4$)
\begin{equation}
R_{\sigma}=\left(\begin{array}{cccc}
1 & 0.4 &  0 & 0 \\
 0.4 & 0 & 0 & 0 \\
0 & 0 & 0 & 0 \\
-0.4 & 0 &  0 & 0.2
\end{array}\right).
\end{equation}
A numerical calculation of its discord, based on the use of orthogonal projectors \cite{povm}, gives ${\cal D}_{A:B}\simeq 2.6*10^{-2}$, while the geometric measure of discord would give approximately $10^{-2}$ \cite{dakic2}.
Despite the fact that $L_{R_{\sigma}}=3$, and then ${\cal Q}>0$, this state would give ${\cal F}=0$ in a RSP protocol.

\section{Conclusions}\label{V} 
 After the discovery that quantum advantages are possible even in the absence of entanglement (the DQC1 protocol), various efforts have been made to find which is the ultimate resource. Quantum discord has been shown to be present in all the known tasks where quantum advantages can be obtained with factorized states.
 On the other hand, one of the most critical and criticized aspects of discord is that it can be generated by local dephasing of a classically correlated state, given that it is hard to assume as a non local resource something that could have been created locally. 

Here we have discussed the role played by quantum discord in a remote state preparation protocol. We have found that,  with no restriction to the form of the initial state used as a channel, the presence of bona fide quantum discord is neither necessary nor sufficient for the protocol to be efficient. The result was supported by two explicit examples. While the presence of fully quantum states that do not give any advantage could be interpreted supposing that some other quantifier of quantum correlations could better characterize the process, the fact that states with locally induced discord play a constructive role probably means that, as far as the specific protocol proposed in Ref. \cite{dakic} is concerned, the function rendered by quantum correlations is far from being understood. As suggested in Ref. \cite{tufarelli},  the link between remote-state preparation and measures of discord, might be due to a nonoptimized version of the
protocol used.
 In any case,  the fact that there are both bona fide quantum states that are of no help in RSP and useful states whose correlations have been generated locally shows that a complete and general operational meaning of quantum discord related to RSP is still missing.

 \subsection*{ \textbf{Acknowledgment}}
  The author acknowledges financial support from Compagnia di San Paolo.

\end{document}